\documentclass[apjl]{emulateapj}
\usepackage{natbib}
\shorttitle{Observations of the Optical Transient in NGC 300 with AKARI/IRC}
\shortauthors{Ohsawa, R., et al.}

\begin{document}

\title{Observations of the Optical Transient in NGC 300 with AKARI/IRC: Possibilities of Asymmetric Dust Formation}
\author{R. Ohsawa\altaffilmark{1},
I. Sakon\altaffilmark{1},
T. Onaka\altaffilmark{1},
M. Tanaka\altaffilmark{2},
T. Moriya\altaffilmark{2,1,3},
T. Nozawa\altaffilmark{2},
K. Maeda\altaffilmark{2},
K. Nomoto\altaffilmark{2},
N. Tominaga\altaffilmark{4,2},
F. Usui\altaffilmark{5},
H. Matsuhara\altaffilmark{5},
T. Nakagawa\altaffilmark{5},
and
H. Murakami\altaffilmark{5}
}
\altaffiltext{1}{Department of Astronomy, Graduate School of Science, University of Tokyo, 7-3-1 Hongo, Bunkyo-ku, Tokyo 113-0033, Japan; ohsawa@astron.s.u-tokyo.ac.jp}
\altaffiltext{2}{Institute for the Physics and Mathematics of the Universe, University of Tokyo, 5-1-5 Kashiwanoha, Kashiwa Chiba 277-8583, Japan}
\altaffiltext{3}{Research Center for the Early Universe, Graduate School of Science, University of Tokyo, 7-3-1 Hongo, Bunkyo-ku, Tokyo, Japan}
\altaffiltext{4}{Department of Physics, Faculty of Science and Engineering, University of Konan, 8-9-1 Okamoto, Higashinada-ku, Kobe, Hyogo 658-8501, Japan}
\altaffiltext{5}{Institute of Space and Astronautical Science, Japan Aerospace Exploration Agency, 3-1-1 Yoshinodai, Chuo-ku, Sagamihara, Kanagawa 252-5210, Japan}

\begin{abstract}
  We present the results of near-infrared (NIR) multi-epoch observations of the optical transient in the nearby galaxy NGC300 (NGC300-OT) at $398$ and $582$ days after the discovery with the Infrared Camera (IRC) onboard AKARI. NIR spectra ($2$--$5\,{\rm \mu m}$) of NGC300-OT were obtained for the first time. They show no prominent emission nor absorption features, but are dominated by continuum thermal emission from the dust around NGC300-OT. NIR images were taken in the $2.4$, $3.2$, and $4.1\,{\rm \mu m}$ bands. The spectral energy distributions (SED) of NGC300-OT indicate the dust temperature of $810\pm14\,{\rm K}$ at $398$ days and $670\pm12\,{\rm K}$ at $582$ days. We attribute the observed NIR emission to the thermal emission from dust grains formed in the ejecta of NGC300-OT. The multi-epoch observations enable us to estimate the dust optical depth as $\gtrsim 12$ at $398$ days and $\gtrsim 6$ at $582$ days at $2.4\,{\rm \mu m}$, by assuming an isothermal dust cloud. The observed NIR emission must be optically thick, unless the amount of dust grains increases with time.  Little extinction at visible wavelengths reported in earlier observations suggests that the dust cloud around NGC300-OT should be distributed inhomogeneously so as to not screen the radiation from the ejecta gas and the central star. The present results suggest the dust grains are not formed in spherically symmetric geometry, but rather in a torus, a bipolar outflow, or clumpy cloudlets.
\end{abstract}

\section{Introduction}
Massive stars are expected to play an important role in the interstellar dust budget in young dwarf galaxies as well as in galaxies in the early universe because of their relatively short evolution lifetime \citep[e.g.,][]{dw07,mo03}. Recently, several eruptive objects with intermediate maximum absolute luminosities between classical nova eruptions and supernova (SN) explosions have been discovered. They are termed as SN impostors \citep{va00} and suggested to be a super-outburst of a luminous blue variable (LBV)-like event rather than a complete disruption of the progenitor. SN impostors provide us with useful opportunities to investigate the effect of an eruptive outburst on the yield of circumstellar dust around the evolved massive stars before core collapse \citep{sa09,ma08}. Among SN impostors, SN 2008S in NGC 6946 \citep{ar08} and the optical transient in NGC 300 \citep[][hereafter NGC300-OT]{mo08} are peculiar ones. One of the outstanding characteristics of them is that their progenitors are deeply embedded in circumstellar dust shells. \citet{kh10} carried out a systematic mid-infrared photometric search for candidate objects which are analogous to the progenitors of SN 2008S and NGC300-OT in four nearby galaxies M33, NGC 300, M 81, and NGC 6946 with \textit{Spitzer}/IRAC. They found dozens of extreme asymptotic giant branch (EAGB) objects similar to SN 2008S and NGC300-OT progenitors, but none of them are brighter than the NGC300-OT progenitor and redder than the SN 2008S progenitor. They point out that the rarity of such extremely red and luminous progenitors can be interpreted as the short duration of the dust-obscured phase with rapid mass loss rates prior to some kind of explosion event in the evolution of massive stars \citep{th09,pu09}.

NGC300-OT was discovered on 2008 May 14 in the nearby spiral galaxy NGC300 \citep{mo08}, the distance of which is $1.9\,{\rm Mpc}$. NGC300-OT has shown an intermediate peak absolute magnitude of $M_{\rm bol} \sim -11.8\, {\rm mag.}$ with an optical spectrum well reproduced by an F-type supergiant photosphere with emission lines of hydrogen, \ion{Ca}{2}, and [\ion{Ca}{2}], similar to typical spectra of low-luminosity SN imposters \citep{bo09,be09}. The progenitor of NGC300-OT was not detected in \textit{Hubble Space Telescope}(\textit{HST}) archive images \citep{bo09}, but was identified in \textit{Spitzer} archival images \citep{pr08}, which suggest the dust-enshrouded nature of the progenitor. Several interpretations have been proposed for the nature of NGC300-OT: a heavily dust-enshrouded OH/IR star of $10$--$15M_{\odot}$ \citep{bo09}, a dust enshrouded star with a luminosity of about $6\times 10^{4}L_{\odot}$ indicative of a $10$--$20M_{\odot}$\citep{be09}, or a massive ($M\sim6$--$10M_{\odot}$) carbon-rich asymptotic giant branch (AGB) star or a post-AGB star \citep{pr09}. The validity of those mass ranges is confirmed by \citet{go09} based on the method that derives the star formation history of a transient's host stellar population. \citet{pr09} also pointed out the presence of newly formed dust of $1500\,{\rm K}$ in addition to the pre-existing circumstellar dust of $\sim 3 \times 10^{-4} M_{\odot}$ of $400\,{\rm K}$ based on the near- to mid-infrared spectral energy distribution (SED) at 93 days after the discovery. Further near-infrared (NIR) observations at later epochs are crucial to investigate the properties of newly-formed dust grains at the outburst.

In this paper, we present our results of NIR observations of NGC300-OT on the 398th and the 582nd days with the Infrared Camera\,(IRC) onboard AKARI \citep{on07}. NIR spectroscopy in $2$--$5\,{\rm \mu m}$ and imaging in three filter bands of \textit{N}2\,($2.4\micron$), \textit{N}3\,($3.2\micron$) and \textit{N}4\,($4.1\micron$) were carried out with the IRC. We focus on the time evolution of the emission from newly formed dust and investigate the dust formation process around NGC300-OT during its eruptive outburst.

\section{Observations and Data Reduction}
The first epoch datasets were obtained on 2009 June 16 corresponding to the epoch of 398 days after the discovery as part of the Director's Time (DT).  They consist of imaging observations in three filter bands centered at $2.4\,\micron$ (\textit{N}2), $3.2\,\micron$ (\textit{N}3), and $4.1\,\micron$ (\textit{N}4) \citep{on07} of the pointing ID  5200806.1 and 5200806.2, and spectroscopic observations with the NIR prism (NP; $\Delta\lambda/\lambda \sim 20$) for $2$--$5\,\micron$ \citep{oh07} of the pointing ID 5200805.1 and 5200807.1. The second-epoch data sets were obtained on 2009 December 17 corresponding to the epoch of 582 days after the discovery as part of the Mission Program (MP) ``Interstellar Medium in our Galaxy and Nearby Galaxies'' \citep[ISMGN;][]{ka09}.  They include imaging observations in \textit{N}2, \textit{N}3, and \textit{N}4 bands (pointing ID 1422250.1 and 1422250.2) and spectroscopic observations with the NIR grism (NG; $\Delta\lambda/\lambda \sim 100$) for $2.5$--$5\,\micron$ \citep{oh07} (pointing ID 1422249.1 and 1422249.2). 

The imaging data were reduced with the AKARI/IRC Imaging Toolkit for Phase 3 data version 20081015\footnotemark[1] and the flux was measured with {\tt{phot}} command in the {\tt{apphot}} package of {\tt{iraf}}\footnotemark[2]. The spectroscopic data were reduced with the AKARI/IRC spectroscopy toolkit for phase 3 data version 20090211. The near infrared spectra of NGC300-OT were taken in the slitless spectroscopy mode. Therefore, they are blended with the spectra of diffuse emission from the host galaxy NGC 300. We carefully subtracted the contribution of diffuse emission from NGC 300 by interpolating the signals around NGC300-OT. The value of a bad pixel was replaced by the median value of the adjoining pixels. The NG spectrum at 582 days was smoothed by a $3$ pixel running mean to increase the signal-to-noise ratio without degrading the spectral resolution.
\footnotetext[1]{\tiny{http://www.ir.isas.jaxa.jp/ASTRO-F/Observation/DataReduction/IRC/}}
\footnotetext[2]{\tiny{http://www.iraf.net/}}

\section{Results and Discussion}
\subsection{Spectral Data}
\begin{figure*}[tb]
  \epsscale{.6}
  \plotone{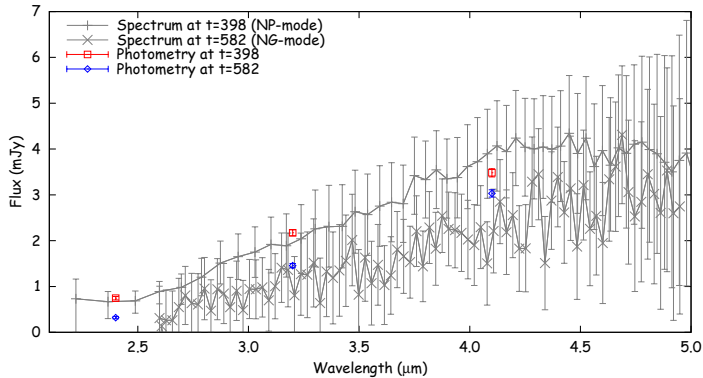}
  \caption{
    NIR spectrum of NGC300-OT. The thicker line with the pluses shows the spectrum at $398$ days taken with the prism ($\lambda/\Delta\lambda {\sim} 20$), the line with the crosses shows the spectrum at $582$ days taken with the grism ($\lambda/\Delta\lambda{\sim}100$), the squares show the fluxes obtained from the NIR images at $398$ days, and the diamonds show the fluxes derived from the NIR images at $582$ days.
    \label{fig:nirspectrum}}
\end{figure*}

Figure \ref{fig:nirspectrum} shows the obtained NIR spectra of NGC300-OT at $398$ and $582$ days. No significant contribution is recognized from the polycyclic aromatic hydrocarbon (PAH) features around $3.3\,\micron$, hydrogen recombination lines of Brackett $\alpha$ at $4.05\,\micron$, and Brackett $\beta$ at $2.63\,\micron$, nor forbidden lines from ionized gas (e.g. [\ion{Mg}{4}] at $4.49\,\micron$, [\ion{Ca}{4}] at $3.21\,\micron$ and [\ion{Ca}{5}] at $4.15\,\micron$) in the spectra. Therefore, the  NIR spectra of NGC300-OT at both epochs are dominated by hot dust continuum emission.
\subsection{Imaging Data}
The results of the photometry are listed in Table \ref{tab:sedofngc300ot}. \citet{pr10} have reported that the $3.6\,\micron$ flux at $587$ days is $2.2\,{\rm mJy}$, which is in agreement with our results at $582$ days. Compared to the fluxes at $398$ days, the fluxes at $582$ days had declined by 56\% in the \textit{N}2, 35\% in the \textit{N}3 and, 15\% in the \textit{N}4 bands.

Since the noise level in the spectroscopic mode at $4.05\,\micron$ is about $3\,{\rm mJy}$ ($3\sigma$) and the estimated spectral resolution is $0.2\,\micron$ for NP and $0.04\,\micron$ for NG, an upper limit of the intensity of Brackett $\alpha$, which is expected to be the strongest among the hydrogen recombination lines in $2$--$5\,\micron$, is estimated as $5.1\times 10^{-17}\,{\rm W\,m^{-2}}$ and $7.8\times 10^{-18}\,{\rm W\,m^{-2}}$ for $398$ and $582$ days, respectively. Thus, the contribution of Brackett $\alpha$ on the \textit{N}4 band is supposed to be negligible.

\begin{table}[bt]
  \centering
  \caption{Photometric Data of NGC300-OT\label{tab:sedofngc300ot}}
  \begin{tabular}{ccc}
    \hline \hline

    \multicolumn{1}{c}{$\lambda~({\rm \mu m})$} &
    \multicolumn{1}{c}{$t{=}398$ days$({\rm mJy})$} &
    \multicolumn{1}{c}{$t{=}582$ days$({\rm mJy})$} \\
    \hline

    $2.4$ & $0.75\pm0.04$ & $0.33\pm0.02$ \\
    $3.2$ & $2.22\pm0.07$ & $1.45\pm0.05$ \\
    $4.1$ & $3.50\pm0.09$ & $2.97\pm0.08$ \\
    \hline
  \end{tabular}
\end{table}

\section{Discussion}
\subsection{Contributions from Two Dust Components}
According to \citet{pr09}, the SED of NGC300-OT at $96$ days is well explained by three blackbody components of temperatures of ${\sim}4000\,{\rm K}$, ${\sim}1500\,{\rm K}$, and ${\sim}480\,{\rm K}$. The component of ${\sim}4000\,{\rm K}$ is assumed to come from the photosphere. The component of ${\sim}480\,{\rm K}$ (hereafter, the warm dust) is attributed to the dust produced in the AGB/post-AGB phase. While the origin of the component of ${\sim}1500\,{\rm K}$ (hereafter, the hot dust) remains to be identified, it is most likely to be attributable to the newly formed dust in the outburst ejecta \citep{pr09,pr10}. If these grains are heated by the radiation from the central star and are in radiative equilibrium, the temperature of the dust should follow the relation of $T_d \propto L^{\frac{1}{4}}$, where $T_d$ is the dust temperature and $L$ is the luminosity of the central star.  According to the light curve by \citet{bo09}, the luminosity of NGC300-OT in the \textit{V}-band had fallen by about $3\,{\rm mag}$ from $96$ to ${\sim}200$ days. The dust temperatures at ${\sim}200$ days should be about a half of that at $96$ days, or the warm dust temperature at ${\sim}200$ days should be ${\sim}240\,{\rm K}$.  We estimate that the contribution to the NIR emission from the warm dust decreases from that at $200$ days and should be at most 1\% of the observed flux at $398$ days, which is lower than the uncertainties in the photometry. Thus, we can attribute the observed change of the SED entirely to the hot dust, which comes from dust grains that were formed in the ejecta.
\subsection{Optical Depth of The Dust Cloud}
\begin{figure*}[tb]
  \epsscale{.75}
  \plotone{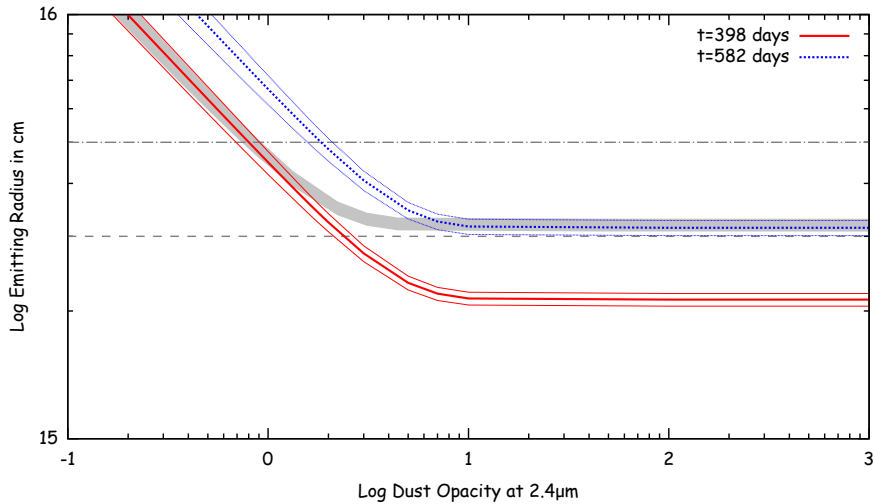}
  \caption{
    Relation between the dust optical depth at $2.4\,\micron$ ($\tau_{\nu}$) and the emitting radius ($R$). The red solid line shows the relation at $t{=}398$ days, while the blue dotted line shows that at $t{=}582$ days. The range of uncertainties is indicated by the thin lines. The dashed and dot-dashed lines are upper limits of the emitting radius, respectively, at $t{=}398$ days ($R=3\times10^{15}\,{\rm cm}$) and at $t{=}582$ days ($R=5\times10^{15}\,{\rm cm}$), corresponding to a shell expansion velocity of $1000\,{\rm km\,s^{-1}}$. Assuming that the dust mass is constant and the dust cloud expands at a constant velocity, the $\tau_{\nu}$ $R$ line at $t{=}398$ days is predicted to evolve to the gray shaded region at $t{=}582$ days (see the text). 
  \label{fig:relopa}}
\end{figure*}

To investigate the change in the properties of the dust, we assume a simple model of an isothermal dust cloud and fit the SED derived from the NIR imaging data by $f_\nu \propto \left(1-e^{-\tau_{\nu}}\right)B_\nu(T_d)$, where $\tau_{\nu}$ is the dust optical depth, and $B_\nu(T_d)$ is the Planck function. We assume that $\tau_{\nu}$ is proportional to $1/\lambda$. We derive dust temperatures for both optically thin ($\tau_{\nu}{\ll}1$) and thick ($\tau_{\nu}{\gg}1$) case.  The dust temperatures and the total luminosities are listed in Table \ref{tab:sedresult}. Based on this simple model, the dust emitting radius $R$ can be given by
\begin{equation}
  \label{eq:emitting}
  R = \left[\frac{D^2f_{\nu}}{\pi\left(1-e^{-\tau_{\nu}}\right)B_\nu(T_d)}\right]^{\onehalf},
\end{equation}
where $D$ is the distance to NGC300-OT from the observer and $f_\nu$ is the observed flux. In Figure \ref{fig:relopa}, $R$ is plotted against $\tau_{\nu}$ at $2.4\,\micron$.  According to \citet{be09}, an upper limit of the expansion velocity of the ejecta is about $1000\,{\rm km\,s^{-1}}$. If the ejecta expands with this velocity, $R$ must be smaller than ${\sim} 3\times 10^{15}\,{\rm cm}$ (the dashed line in Figure \ref{fig:relopa}) at $398$ days and then ${\sim} 5\times10^{15}\,{\rm cm}$ (the dot-dashed line) at $582$ days. Assuming that the dust mass did not change between $398$ and $582$ days, the dust opacity should be proportional to $R^{-2}$. If the dust cloud expands at a constant velocity, the ratio of the radius at $582$ days to that at $398$ days must be equal to $582/398 \sim 1.5$.  Therefore, the $\tau_{\nu}$ $R$ line at $398$ days is predicted to evolve to the gray shaded region in Figure \ref{fig:relopa}. Only the region where the predicted line overlaps with the observed line at $582$ days gives consistent models both at 398 and 582 days. From Figure \ref{fig:relopa}, a lower limit of $\tau_{\nu}$ ($2.4\,\micron$) is estimated to be ${\sim}12$ at $398$ days and ${\sim}6$ at $582$ days. Therefore, the dust cloud around NGC300-OT is optically thick at both epochs. It is consistent with the featureless spectra shown in Figure \ref{fig:nirspectrum}. Assuming the constant density, we can estimate a lower limit of the dust mass as $M_d = (4\pi/3)R^2\tau_\nu\kappa_\nu{}^{-1} \sim 10^{-5}\,M_\odot$, where $\kappa_\nu$ is the dust mass absorption coefficient at $2.4\,\micron$.  The absorption coefficient $\kappa_\nu$ is estimated as $(3/4)2\pi\lambda^{-1}\rho^{-1}\sim10^4\,{\rm cm^2\,g^{-1}}$, where $\rho$ is the mass density of the dust grains, ${\sim}\,2\,{\rm g\,cm^{-3}}$, assuming carbon dust \citep{pr09}. In case that the density varies as $r^{-2}$, we need to assume the inner shell radius for the dust mass estimation. If the inner shell radius is within a half or a quarter of $R$, the estimated dust mass will not change more than ${\pm}40\,\%$ of that in the constant density case.

\begin{table}[bt]
  \centering
  \caption{The results of SED fit\label{tab:sedresult}}
  \begin{tabular}{lcc}
    \hline \hline
    \multicolumn{1}{c}{} &
    \multicolumn{1}{c}{$398$ days} &
    \multicolumn{1}{c}{$582$ days}
    \\ \hline

    \multicolumn{3}{l}{\footnotesize for $\tau \gg 1$}
    \\ \hline
    $T_d\,~({\rm K})$        & $813\pm14$  & $668\pm12$  \\
    $L_d\,({10^5\!L_\odot})$  & $4.0\pm0.4$ & $3.4\pm0.3$
    \\ \hline

    \multicolumn{3}{l}{\footnotesize for $\tau \ll 1$}
    \\ \hline
    $T_d\,~({\rm K})$        & $678\pm10$  & $573\pm12$  \\
    $L_d\,({10^5\!L_\odot})$  & $4.0\pm0.5$ & $3.3\pm0.4$
    \\ \hline

  \end{tabular}
\end{table}

If the dust cloud is optically thick, there must be a temperature gradient. Thus an isothermal model is a crude approximation, particularly for the optically thick case. To examine how secure the conclusion of the optically thick emission is, we consider an optically-thin cloud in some more detail. The total dust mass can be estimated for an optically thin case using Equation (4) in \citet{dw83}:
\begin{equation}
  \label{eq:dweketal}
  M_d \simeq 2\times 10^{-6}\left(\frac{T_d}{1000\,{\rm K}}\right)^{-5}\left(\frac{L_d}{10^6L_\odot}\right)\,M_\odot,
\end{equation}
where $L_d$ is the total dust luminosity in $L_\odot$. Using the temperature and the luminosity in Table \ref{tab:sedresult} for $\tau \ll 1$, we obtain $M_d \sim 5.1\times 10^{-6}\,M_\odot$ at $398$ days and $M_d \sim 1.0\times10^{-5}\,M_\odot$ at $582$ days. Thus, if the emission is optically thin, the dust mass must be increased by a factor of ${~}2$. Theoretical investigations suggest that the nucleation and grain growth will cease in a relatively short time right after the onset of nucleation \citep{ya77,dr77}. Therefore, we conclude that optically thin models cannot account for the observations.

Since the ratio of the opacity at the \textit{V}-band to that at $2.4\,{\rm \mu m}$ is about $5$, the optically-thick dust cloud predicts strong extinction at visible wavelengths. \citet{be09} reported $E(B{-}V) = 0.05\pm0.05\,{\rm mag.}$ at $121$ days, which suggests that the extinction by the dust cloud is very small or negligible. \citet{pr10} reported that the \textit{R}-band magnitude has significantly faded to $23.9\pm 0.2\,{\rm mag.}$ at $585$ days and they surmise that NGC300-OT is likely becoming self-enshrouded. An upper limit of the extinction at the \textit{R}-band at $585$ days can be estimated as $6\,{\rm mag.}$, if we simply assume the \textit{R}-band magnitude at $585$ days is the same as that at $120$ days. Although it may be true that part of the stellar radiation is attenuated, it is too small to account for the extinction caused by the optically thick dust cloud, more than ${\sim}20\,{\rm mag.}$, in the preceding discussion. The apparent contradiction suggests that the dust around NGC300-OT is distributed not uniformly, but in spherically asymmetrical geometry, such as a torus, bipolar, or clumpy form, so that the dust cloud does not completely screen the radiation from the ejecta gas and the central star. The existence of a significantly asymmetrical dust cloud was also proposed by \citet{pa10} and \citet{be09}. The present results also support the existence of an asymmetrical dust cloud.

\section{Summary}
We present the results of the NIR observations on NGC300-OT at $398$ and $582$ days after the discovery with the IRC onboard AKARI. NIR spectra ($2$--$5\,{\rm \mu m}$) are obtained for the first time. They show no prominent emission or absorption features but are dominated by continuum radiation of the dust around NGC300-OT. NIR photometric data were taken in \textit{N}2, \textit{N}3, and \textit{N}4 bands. The SED of NGC300-OT indicates a dust temperature of $810\pm14\,{\rm K}$ at $398$ days and $670\pm12\,{\rm K}$ at $582$ days. We estimate the dust optical depth as $\gtrsim 12$ on day $398$ and $\gtrsim 6$ on day $582$, assuming an isothermal dust cloud. Although the present model is very crude, optically thin models cannot account for the observations at day 398 and 582 consistently. The large optical depths at NIR suggest a large extinction at visible wavelengths, while little extinction is reported based on earlier observations ($E(B-V) = 0.05 \pm 0.05$).  Those circumstances suggest that the dust cloud around NGC300-OT should be distributed asymmetrically so as to not screen the radiation from the ejecta gas and the central star. The present results suggest that the dust formation occurs in NGC300-OT in the form of a torus, a bipolar outflow, or clumpy cloudlets.

\acknowledgments
This work is based on observation of AKARI, a JAXA project with the participation of ESA. We thank all the members of the AKARI project, particularly those who have engaged in the observation planning and the satellite operation during the performance verification phase, for their continuous help and support. We would also express our gratitude to the AKARI data reduction team for their extensive work in developing data analysis pipelines. This work is supported in part by Grants-in-Aid for Scientific Research from the JSPS and in part by World Premier International Research Center Initiative (WPI Initiative), MEXT, Japan.

Facilities: 
\facility{AKARI(IRC)}

\end{document}